\documentclass[twocolumn,showpacs,preprintnumbers,amsmath,amssymb,superscriptaddress]{revtex4}

\usepackage{graphicx}
\usepackage{dcolumn}
\usepackage{bm}

\begin{document}
\title{On the RKKY range function of a one dimensional non interacting electron gas}
\author{Gabriele F. Giuliani}
\affiliation{Department of Physics, Purdue University, West
Lafayette, IN 47907}
\email{gfg@physics.purdue.edu,tdatta@physics.purdue.edu}
\author{Giovanni Vignale}
\affiliation{Department of Physics, University of Missouri, Columbia,
Missouri 65211} \email{vignaleg@missouri.edu}
\author{Trinanjan Datta}
\affiliation{Department of Physics, Purdue University, West
Lafayette, IN 47907}
\date{\today}
\begin{abstract}
{\flushleft} We show that the pitfalls encountered in earlier
calculations of the RKKY range function for a non interacting one
dimensional electron gas at zero temperature can be unraveled
and successfully dealt with through a proper handling of the impurity potential.
\end{abstract}

\pacs{71.10.Ca}

\maketitle
The apparently straightforward evaluation of the
Ruderman-Kittel-Kasuya-Yosida (RKKY) range function, or more
generally, of the linear density modulation $\delta n (x)$ induced
at zero temperature in a non interacting one dimensional electron
gas by a localized static impurity modeled with a delta function
potential has proven surprisingly troublesome. While the original 
calculation gave an incorrect answer\cite{Kittel}, more recent 
investigations appear to suggest that only a certain procedure, 
known to lead to a physically sensible answer, should be
employed\cite{Yafet}.

The original and most popular procedure is based on the standard
theory of linear response\cite{TheBook}. If the impurity potential
is assumed to be of the form $U(x) = \frac{\hbar^2 u}{2m} \delta
(x)$ (where $u$ is a suitable wave vector) then one has
\begin{equation}
\label{deltanx_lrt} \delta n(x) ~ = ~  \frac{\hbar^2 u}{2mL}
\sum_{q}\chi_{0}(q,0)e^{iqx} ~,
\end{equation}
where $L$ is the system length and $\chi_{0}(q,0)$ is the static 
Lindhard response function in one dimension given by
\begin{equation}
\label{1DLindhard_implicit} \chi_{0}(q,0) ~=~ \frac{2}{L} {\cal P}
\sum_{k} \frac{n_{k}-n_{k+q}}{\epsilon_{k+q} -\epsilon_{k}} ~,
\end{equation}
where the notation ${\cal P}$ stipulates that the principal part of the 
sum must be taken.
Substituting (\ref{1DLindhard_implicit}) in (\ref{deltanx_lrt}) and using the
zero temperature occupation numbers immediately leads to the formula
\begin{equation}
\label{generalformula} \delta n(x) ~=~ \frac{u}{2\pi^{2}}
\int_{-\infty}^{\infty}dq ~e^{iqx} \int_{-k_{F}}^{k_{F}}dk
\left(\frac{1}{2kq+q^{2}}-\frac{1}{2kq-q^{2}}\right) .
\end{equation}
The interpretation and the handling of this expression rest at the
origin of the problem at hand.

If one simply proceeds to explicitly evaluate the integral over
$k$ in Eq.~(\ref{generalformula}), or, which is the same, that of
(\ref{1DLindhard_implicit}), the result is well known and is given
by
\begin{equation}
\chi_{0}(q,0) ~=~ \frac{2m}{\pi \hbar^{2}q} \ln \left
|\frac{2k_{F}+q}{2k_{F}-q} \right| ~.
\end{equation}

At this point the second integral over $q$ can be readily evaluated
leading to
\begin{equation}
\label{deltanx-right} \delta n(x) ~ = ~ -\frac{u}{\pi} si (2k_F x) ~,
\end{equation}
where the sine integral function appears\cite{GR}. This result
provides the correct answer to the problem. In particular the large
distance
behavior turns out to be
\begin{equation}
\label{Friedel-limit} \delta n(x) ~ \simeq ~  - \frac{u}{\pi}
\frac{\cos(2k_Fx) }{ x} ~,
\end{equation}
an expression displaying the expected decay and Friedel
oscillations\cite{Friedel}. One should notice at this point that, in
view of the singular behavior of the integrand at the origin,
however tempting, the order of the $k$ and $q$ integrations in
Eq.~(\ref{generalformula}) cannot be freely exchanged\cite{Yafet}.
Doing so leads to the manifestly unphysical answer\cite{Kittel}
\begin{equation}
\label{wrongnx} \delta n(x) ~ = ~ - \frac{u}{\pi}  \left( si (2k_F
x) + \frac{\pi}{2} \right) ~~~(not!) ~.
\end{equation}
This pitfall is unique to the one dimensional case for in two and
three dimensions formulas equivalent to (\ref{generalformula}) can
be derived and safely evaluated by exchanging at will the order of
the intervening integrations.

An alternative, appealing and equally physically valid procedure to
obtain $\delta n (x)$ is in the case offered by extracting the
leading linear term of the formula
\begin{equation}
\label{deltanx-secondway} n(x) - n_0 ~=~ 2\sum_{|\vec{k}| \leq
k_{F}} (|\psi_{k}(x)|^{2}-|\psi_{0}(x)|^{2}) ~,
\end{equation}
valid for single Slater determinant states. Now, if the usual first
order perturbation theory result for $\psi_{k}(x)$ is used in
(\ref{deltanx-secondway}), one is immediately led to an equation
that differs from (\ref{generalformula}) merely for the exchange of
the order of the quadratures\cite{Kittel}. Since, as observed above,
such an expression does lead to an unphysical result, it has been
suggested that this route is unphysical and should be
avoided\cite{Yafet}. We find that this conclusion is unwarranted
for, as we show next, the difficulty lies here with the use of
perturbation theory, which is invalid, and not with
Eq.~(\ref{deltanx-secondway}) per se.

To prove our assertion we observe that the exact delocalized eigenstates 
of the Schr\"odinger equation
\begin{equation}
\label{Schroedinger_equation} \frac{\hbar^{2}}{2m}\left\{-
\frac{\partial^{2}}{\partial x^{2} } + u \delta (x) \right\}\psi(x)
~= ~E\psi(x) ~,
\end{equation}
can be written as a superposition of the two following (normalized) 
scattering states 
\begin{equation}
\psi_{k +} (x) ~=~ \left\{
\begin{array}{lll}
 \frac{e^{i kx}}{\sqrt{L}} + \frac{u}{2ik  - u} \frac{e^{ - i kx}}{\sqrt{L}}
&,& x < 0  \\
\frac{2ik}{2ik  - u} \frac{e^{i kx}}{\sqrt{L}} &,& x > 0
\\ \end{array}  \right.
\end{equation}
and
\begin{equation}
\psi_{k -} (x) ~=~ \left\{
\begin{array}{lll}
 \frac{2ik}{2ik  - u} \frac{e^{- i kx}}{\sqrt{L}}
&,& x < 0  \\
\frac{e^{-i kx}}{\sqrt{L}} + \frac{u}{2ik  - u} \frac{e^{i kx}}{\sqrt{L}}
&,& x > 0,  \\
\end{array}  \right.
\end{equation}
where $k$ is limited to positive values only. By taking the modulus square of the
sum of $\psi_{k +} (x)$ and $\psi_{k -} (x)$ and by summing over $k >0$ in
(\ref{deltanx-secondway}) one obtains the following exact expression
for the electronic density $n(x)$:
\begin{equation}
\label{exactn} n(x) = \frac{2}{\pi }\int^{k_F}_0 dk \left( 1+
\frac{2uk \sin(2k|x|)}{4k^2+u^2} -\frac{u^{2}
\cos(2k|x|)}{4k^{2}+u^{2}} \right) ~.
\end{equation}

At this point we are left with extracting the linear term in $u$.
This must be done with some care. In particular one must resist the
temptation to simply drop the last term in Eq.~(\ref{exactn}) and,
at the same time, to neglect the $u^{2}$ in the denominator of the
second one. To do so coincides with making use of first order
perturbation theory and leads exactly to the original unphysical
result of Eq.~(\ref{wrongnx}). While it is safe to handle the second
term of (\ref{exactn}) as just described, the third term does
contribute a first order term that can be readily extracted by
making use of the relation
\begin{equation}
\lim_{u \rightarrow 0} \frac{u}{4k^{2}+u^{2}} ~=~ \frac{\pi}{2} \delta(k) ~,
\end{equation}
leading to
\begin{equation}
n(x) \simeq \frac{2k_{F}}{\pi} ~+ ~\frac{u}{\pi}\left(
\int^{k_{F}}_{0}dk \frac{ \sin(2k |x|)}{k} - \frac{\pi}{2} \right) ~,
\end{equation}
a result that can be readily seen to coincide with the correct
answer of Eq.(\ref{deltanx-right}).

We have therefore shown that a proper handling of the impurity
potential allows one to correctly carry out the calculation of the
linear density modulation via either the response function method of
Eq.~(\ref{deltanx_lrt}) or the alternative direct procedure offered
by (\ref{deltanx-secondway})\cite{comment_on_finte_T}.

We conclude by remarking that a proper treatment of the problem of
the effects of a static localized impurity in an interacting one
dimensional electron liquid can be achieved by means of the
Luttinger liquid model. The problem is highly non trivial as, while Friedel-like 
oscillations with an amplitude decay generally ruled by the
interaction strength exist at intermediate distances\cite{TheBook},
for large distances the physics of the phenomenon is non linear in
the impurity potential\cite{EggerandGrabert}. Finally a recent
discussion of the effects of the Coulomb interaction on the Friedel
oscillations of an electron liquid in two and three dimensions can
be found in reference \onlinecite{SimionGFG}.

\begin{acknowledgements}
The authors wish to thank A. W. Overhauser and George E. Simion for
useful conversations.
\end{acknowledgements}


\begin{thebibliography}{8}
\expandafter\ifx\csname natexlab\endcsname\relax\def\natexlab#1{#1}\fi
\expandafter\ifx\csname bibnamefont\endcsname\relax
  \def\bibnamefont#1{#1}\fi
\expandafter\ifx\csname bibfnamefont\endcsname\relax
  \def\bibfnamefont#1{#1}\fi
\expandafter\ifx\csname citenamefont\endcsname\relax
  \def\citenamefont#1{#1}\fi
\expandafter\ifx\csname url\endcsname\relax
  \def\url#1{\texttt{#1}}\fi
\expandafter\ifx\csname urlprefix\endcsname\relax\def\urlprefix{URL }\fi
\providecommand{\bibinfo}[2]{#2}
\providecommand{\eprint}[2][]{\url{#2}}

\bibitem[{\citenamefont{Kittel}(1968)}]{Kittel}
\bibinfo{author}{\bibfnamefont{C.}~\bibnamefont{Kittel}},
  \emph{\bibinfo{title}{in Solid State Physics}}, vol.~\bibinfo{volume}{22}
  (\bibinfo{publisher}{Academic Press, New York}, \bibinfo{year}{1968}),
  \bibinfo{note}{edited by F. Seitz, D. Turnbull, and H. Ehreinreich (see also
  erratum ibidem)}.

\bibitem[{\citenamefont{Yafet}(1987)}]{Yafet}
\bibinfo{author}{\bibfnamefont{Y.}~\bibnamefont{Yafet}},
  \bibinfo{journal}{Phys.\ Rev. B} \textbf{\bibinfo{volume}{36}},
  \bibinfo{pages}{3948} (\bibinfo{year}{1987}).

\bibitem[{\citenamefont{Giuliani and Vignale}(2005)}]{TheBook}
\bibinfo{author}{\bibfnamefont{G.~F.} \bibnamefont{Giuliani}} \bibnamefont{and}
  \bibinfo{author}{\bibfnamefont{G.}~\bibnamefont{Vignale}},
  \emph{\bibinfo{title}{Quantum Theory of the Electron Liquid}}
  (\bibinfo{publisher}{Cambridge University Press, Cambridge},
  \bibinfo{year}{2005}).

\bibitem[{\citenamefont{Gradshteyn and Ryzhik}(1965)}]{GR}
\bibinfo{author}{\bibfnamefont{I.}~\bibnamefont{Gradshteyn}} \bibnamefont{and}
  \bibinfo{author}{\bibfnamefont{I.}~\bibnamefont{Ryzhik}},
  \emph{\bibinfo{title}{Table of Integrals, Series and Products}}
  (\bibinfo{publisher}{Academic Press, San Francisco}, \bibinfo{year}{1965}).

\bibitem[{\citenamefont{Friedel}(1954)}]{Friedel}
\bibinfo{author}{\bibfnamefont{J.}~\bibnamefont{Friedel}},
  \bibinfo{journal}{Advanc. Phys.} \textbf{\bibinfo{volume}{3}},
  \bibinfo{pages}{446} (\bibinfo{year}{1954}).

\bibitem[{com()}]{comment_on_finte_T}
\bibinfo{note}{Of course the correct answer can also be arrived at by
  calculating $\delta n (x)$ at finite temperature and then taking the $T
  \rightarrow 0$ as done for instance in V. Litvinov and V. K. Dugaev, Phys.\
  Rev. B {\bf 58}, 3584 (1998). This sheds no light onto the problem at hand.}

\bibitem[{\citenamefont{Egger and Grabert}(1995)}]{EggerandGrabert}
\bibinfo{author}{\bibfnamefont{R.}~\bibnamefont{Egger}} \bibnamefont{and}
  \bibinfo{author}{\bibfnamefont{H.}~\bibnamefont{Grabert}},
  \bibinfo{journal}{Phys. Rev. Letters} \textbf{\bibinfo{volume}{75}},
  \bibinfo{pages}{3505} (\bibinfo{year}{1995}).

\bibitem[{Sim()}]{SimionGFG}
\bibinfo{note}{G. Simion and G. F. Giuliani, to be published}.

\end{thebibliography}
\end{document}